# Structure and Properties of DNA Molecules Over The Full Range of Biologically Relevant Supercoiling States

Bettotti P.[1*], Visone V.[2*], Lunelli L.[3,4], Perugino G.[2], Ciaramella M.[2§], Valenti A.[2§]

1. Nanoscience Laboratory, Department of Physics, University of Trento, Via Sommarive 14, I-38123 Povo (Trento), Italy.
2. Institute of Biosciences and Bioresources, National Research Council of Italy, Via Pietro Castellino 11, 80131, Napoli, Italy.
3. Laboratory of Biomarker Studies and Structure Analysis for Health, FBK-Fondazione Bruno Kessler, Via Sommarive 18, 38123, Povo, Trento, Italy.
4. Institute of Biophysics, National Research Council of Italy, Trento, Italy.
* These authors contributed equally to this work
§ corresponding authors

**Abstract**

Topology affects physical and biological properties of DNA and impacts fundamental cellular processes, such as gene expression, genome replication, chromosome structure and segregation. In all organisms DNA topology is carefully modulated and the supercoiling degree of defined genome regions may change according to physiological and environmental conditions. Elucidation of structural properties of DNA molecules with different topology may thus help to better understand genome functions. Whereas a number of structural studies have been published on highly negatively supercoiled DNA molecules, only preliminary observations of highly positively supercoiled are available, and a description of DNA structural properties over the full range of supercoiling degree is lacking. Atomic Force Microscopy (AFM) is a powerful tool to study DNA structure at single molecule level. We here report a comprehensive analysis by AFM of DNA plasmid molecules with defined supercoiling degree, covering the full spectrum of biologically relevant topologies, under different observation conditions. Our data, supported by statistical and biochemical analyses, revealed striking differences in the behavior of positive and negative plasmid molecules.

1. **Introduction**

DNA topology is an intrinsic property of DNA molecules, and is controlled by the direct action of DNA topoisomerases [1-8]. These are enzymes essential for proliferation and survival of all cells and are indeed important targets for chemotherapeutic drugs. *In vitro,* DNA topoisomerases are able to induce significant changes of linking number (ΔLk; see the Experimental Section for definition of the topological parameters) of covalently closed DNA molecules, by either removing or introducing supercoils. Each enzyme has its own specificity, and can modify the ΔLk stepwise, leading to production of a range of differently supercoiled molecules.

*In vivo*, genomic DNA is organized in topologically closed domains, whose supercoiling degree is strictly regulated, given the great impact of topology on all DNA activities [9-10]. In organisms living in the range of mesophilic temperatures, DNA is in general negatively supercoiled (–SC ); thermophilic microorganisms, living in environmental niches above 70°C, possess a special DNA topoisomerase called reverse gyrase (RG), which is able to catalyze positive supercoiling [11-13]. Recent results show that RG is essential for growth at 95 °C of a hyperthermophilic species, suggesting that the presence of the enzyme is necessary to maintain the correct DNA supercoiling in organisms living at high temperatures [14].

Positively supercoiled (+SC) DNA accumulates in every organism during DNA transactions, such as replication and transcription [15]; moreover, +SC DNA prevents extensive reversal of replication forks in the presence of replication blocking lesions [16], promotes telomere resolutions [17] and marks centromeres as unique chromosome loci [18]. In addition, condensin and cohesin complexes, which play essential role in chromosome architecture and segregation during mitosis and meiosis, induce positive supercoiling [19-20]. Although current techniques do not allow determination of supercoiling degree *in vivo* in real time, it is likely that the full range of topological degree may occur during the cell life and DNA supercoiling may vary significantly over the cell cycle, in distinct genome locations as well as in response to environmental stimuli. Thus, understanding how DNA is organized over the full range of supercoiling degree may provide crucial insights into the principles that underlie genomic organization and regulation in living organisms.

So far, the structure of –SC plasmids has been investigated by several techniques. Electron (SEM, TEM) and scanning probe microscopy (SPM) are the *de-facto* techniques to visualize DNA at single molecule level and to unveil structural details. All these techniques suffer from possible artifacts during sample preparation, which might influence the measurements outcome [23-26]. Atomic force microscopy (AFM) may in part overcome these limitations, as it allows to control several parameters during plasmids deposition (e.g. buffer ionic strength, temperature, etc.) and to simulate *in-vivo* conditions by performing measurements in liquids. Although many structural studies have been reported on –SC plasmids, little information is available for +SC plasmids. Only recently two papers reported the first observations at single molecule level of +SC DNA molecules, giving important impulse to the field [27-28]. However, further analyses are required to extend these results and fill important gaps. Indeed, Irobaliescva et al used cryo-tomography to analyze DNA minicircles of a few hundred base pairs with defined degrees of supercoiling. Although these molecules are useful models, they can only accommodate a few supercoils, due to their short size. More recently, Li et al analyzed by AFM +SC plasmids and measured different lengths between positive and negative topoisomers, whose ΔLk was however undefined. These measurements rely on few molecules only and their conclusions might be of limited relevance because of the reduced ensemble investigated [29]

In the present work, we present a systematic AFM analysis of DNA plasmids with ΔLk covering a wide range of supercoiling (ranging from –12 to +12). Observations were performed under a number of different conditions, and data are supported by quantitative and biochemical analysis, highlighting significant differences in the behavior of –SC and +SC molecules. The relevance of these results on the biological properties of DNA is discussed.

## 2. Results and Discussion

### 2.1 Effect of immobilization conditions on structural conformation of plasmid molecules with different supercoiling degree observed by AFM

RG is a powerful tool to obtain plasmid molecules with different degree of supercoiling: indeed, by changing reaction parameters (temperature, time, DNA/enzyme ratio and so on) it is possible to modulate the enzyme activity and obtain distinct populations of topological isomers with defined ΔLk (called topoisomers; see Figure. 1A for a scheme of the reaction) [30]. Reactions were set up using the pBluScript plasmid (pBs), purified from *Escherichia coli* cultures in its negatively supercoiled form, and the RG from the archaeon *Sulfolobus solfataricus* [10-12,30]. In a time course experiment, aliquots of products were withdrawn after different incubation times and analysed by two-dimensional gel electrophoresis, enabling separation of –SC and +SC topoisomers (Fig. 1A). Fig. 1B shows supercoiling modification of the pBs plasmid with time, from –SC (time 0) to +SC (300 sec.). The topological state of each topoisomer population was expressed calculating the mean specific linking number (mean σ) obtained from the σ values of the most abundant topoisomers produced in each reaction (see Experimental Procedures). In Fig. 1C the mean σ of each topoisomer population was plotted as a function of time, showing the increase of plasmid supercoiling density during reverse gyrase reaction.

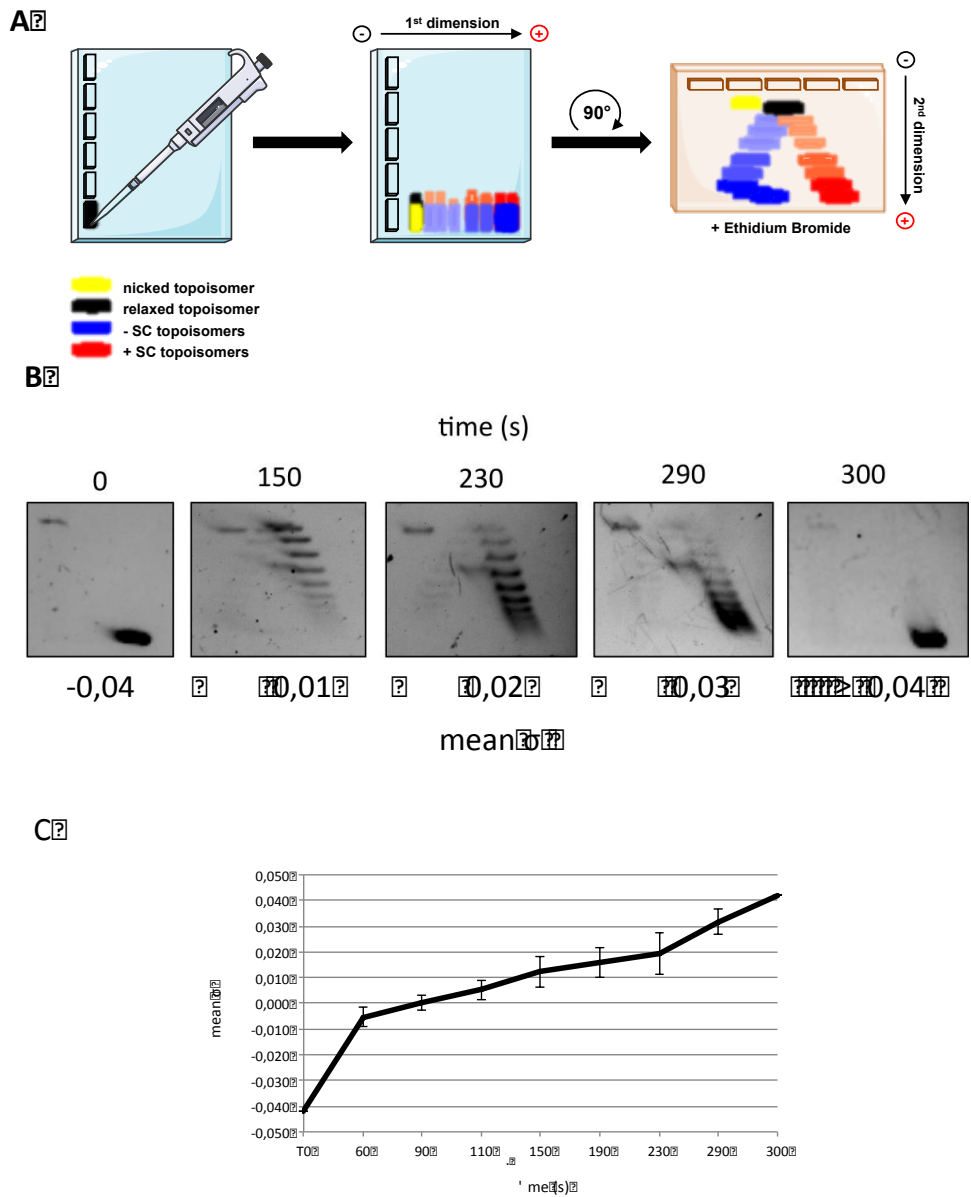

**Figure 1. Reverse gyrase time course reaction analysed by 2D-gel electrophoresis.** A. Schematic representation of 2D-gel migration of –SC and +SC topoisomers. B. –SC pBs DNA was incubated with *S. solfataricus* RG [30] at 90 °C for the time indicated and subjected to 2D electrophoresis. For each time point, the mean σ was determined as described in Experimental Procedures. Representative gel is shown. C. The graph shows increasing of positive supercoiling (expressed as mean σ) with time during RG reaction. The curve is not completely linear, due to the complexity of the reaction, which is affected by a multiplicity of parameters [30, 31]. Data are from three independent experiments.

Among time course products, four plasmid populations with defined mean σ (-0,04; +0,01; +0,03; ≥+0,04) were selected, purified and analysed by AFM in air. Since the only images available so far for positive plasmids were obtained on silanized mica [28], it is not known whether these molecules are affected by the deposition procedure, prompting us to explore different deposition methods. It has been suggested that deposition on mica in the presence of bivalent cations tend to freeze

DNA molecules in their actual configuration, whereas mica treated with silanes allows for the equilibration of molecules during deposition [32]. We tested several procedures and explored a wide range of experimental conditions to find deposition protocols to obtain a mild, yet reproducible, binding of plasmids while preserving the specific features of molecules with different supercoiling degree. We selected appropriate conditions for deposition on mica either in the presence of $MgCl_2$ or after treatment with 3-Amino propyl-trimethoxysilane (APTMS).

When deposited on APTMS-treated mica in the presence of 100 mM NaCl, highly –SC plasmids (σ=-0,04) assumed a plectonemic form with several crossings seen in each molecule and, occasionally, irregularly balled-up or aggregated forms (Fig. 2A). These shapes were already reported for other plasmids having a similar supercoiling degree [33]. In contrast, the populations of topoisomers with an intermediate degree of supercoiling (+0,01< mean σ <+0,03) lost the plectonemic form and showed a reduced number of crossings. Finally, when σ reached the value of +0,04, the molecules adopted again a highly plectonemic conformation (Fig. 2A).

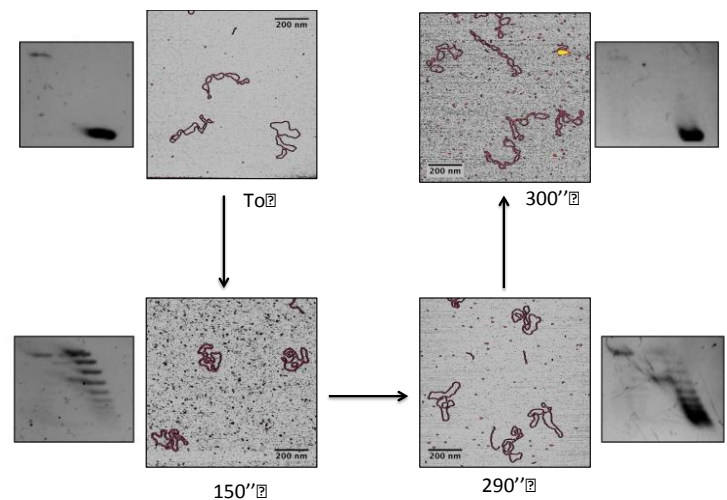

**Figure 2. A.** AFM imaging of distinct topoisomer populations deposited on APTMS-treated mica. Aliquots of DNA were withdrawn during RG reaction at indicated reaction time and analyzed by AFM on APTMS-treated mica. Insets show the migration on 2D gel of each purified population at time 0 (σ=-0,04), after 150" (σ=+0,01), 290" (σ=+0,03) and 300" (σ≥+0,01) of reaction.

Striking different results were obtained when the same samples were deposited in the presence of $MgCl_2$. Highly –SC plasmids adopted a loose geometry, looking like open circular molecules or showing very few crossovers (Fig. 2B), similar to those reported by Y. Lyubchencko [34] A clear transition from relaxed-like to plectonemic structures was observed at σ value of +0,04, showing plectonemic shapes, as observed for the same molecules deposited on APTMS-treated mica. Thus, under these conditions the plectonemic aspect seems a peculiar feature of +SC molecules with a mean σ ≥ +0,04.

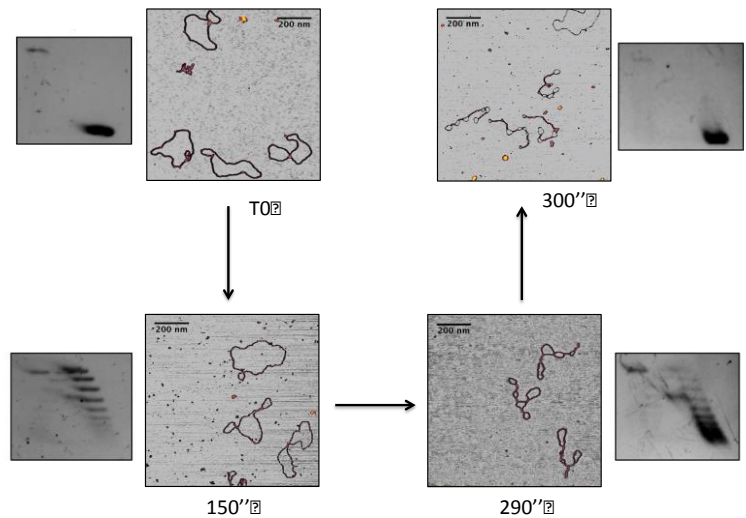

**Figure 2 B.** AFM imaging of distinct topoisomer populations deposited on mica in the presence of MgCl$_2$. Aliquots of DNA were withdrawn during RG reaction at indicated reaction time and analyzed by AFM in the presence of 10 mM MgCl$_2$. . Insets show the migration on 2D gel of each purified population at time 0 (σ=-0,04), after 150" (σ=+0,01), 290" (σ=+0,03) and 300" (σ≥+0,01) of reaction.

Extensive comparison of the shapes of highly +SC molecules on the two surfaces showed that they appear highly homogeneous (Fig. 3), thus suggesting that the shape of these molecules is not heavily affected by the depositions conditions. Moreover, these images show that the end-point product of RG is a homogeneous population comprised of a single, or a few, topoisomer(s) with very close ΔLk.

**A**　　　　　　　　　　　**B**

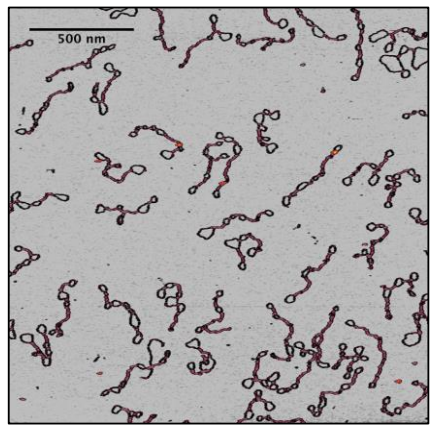
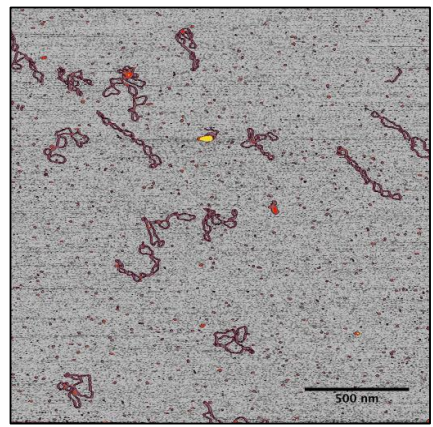

**Mg**　　　　　　　　　　　**APTMS**

**Figure 3.** Comparison of the shapes of highly +SC plasmids (σ ≥+0,04) deposited on mica under different conditions: A. In the presence of 10 mM MgCl$_2$; B. APTMS-treated mica. Representative AFM images are shown.

In order to give statistical significance to our data, we sought to perform quantitative analysis of the structural conformations adopted by the four plasmid populations. As pointed out by Irobaliescva et al. [27], 386 bp long plasmids arrange into a number of different configurations. Because the energy required to supercoil a plasmid is inversely related to its length, longer molecules show even broader range of configurations. Thus, a detailed analysis of the structural conformations may require algorithms currently not available. We developed a numerical routine that traces plasmids from AFM images and extracts a set of geometrical parameters able to discriminate the sign of the supercoiling. While common geometrical descriptors (asphericity, eccentricity, solidity, etc.) were not effective in classifying supercoiling sign (data not shown), we found that the following two parameters quantitatively described the different topoisomer populations:

- density: defined as the ratio between the area occupied by the plasmid and the entire area of the convex hull defining the plasmid;
- Euler number: defined as the number of objects minus the numbers of holes they contain (essentially one object, the plasmid, minus the number of holes contained within it).

We found that higher density and lower Euler number were associated with increased plectonemic character: highly +SC and –SC populations deposited on APTMS-treated mica showed overlapping density and Euler number, consistent with their supercoiled aspect (Fig. 4A); in contrast, both parameters were able to discriminate between the two populations observed in the presence of $MgCl_2$ (Fig. 4B). This fact suggests that –SC and +SC plasmids assume different shapes because of specific interactions with the $Mg^{2+}$ cations. Plasmid populations with intermediate $\Delta Lk$ showed intermediate plectonemic character when deposited on APTMS-treated mica (Fig. 4C), whereas all populations except the highly +SC looked essentially relaxed on $MgCl_2$ (Fig. 4D).

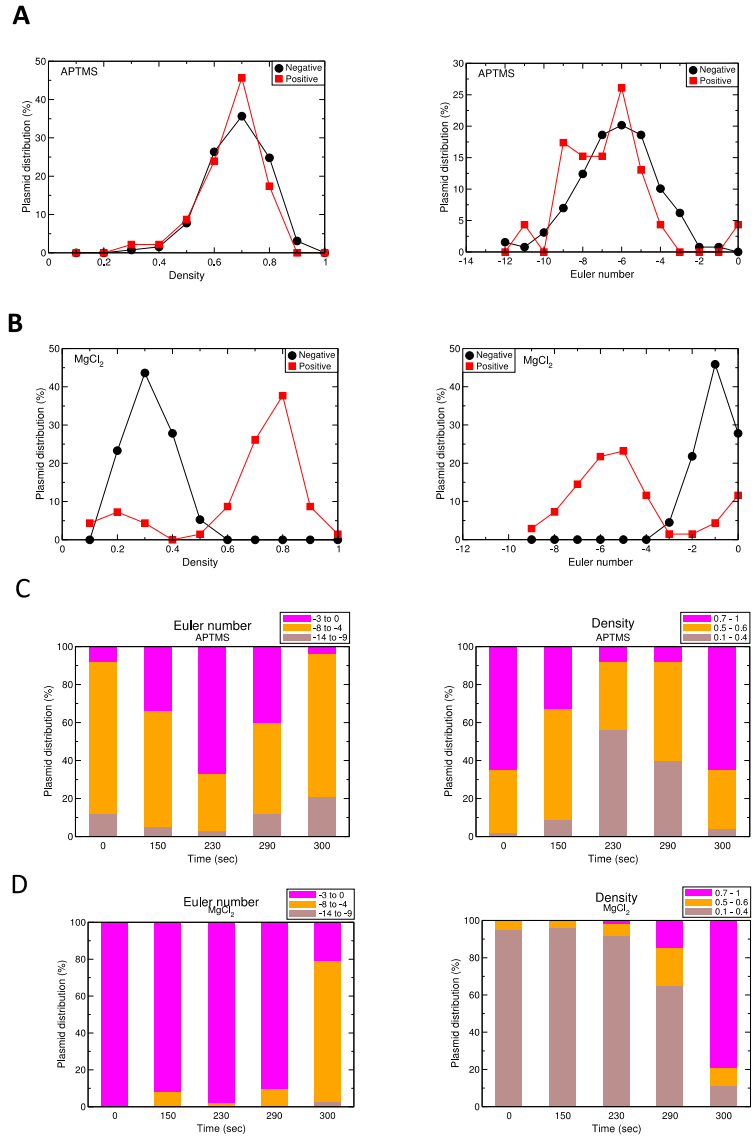

**Figure 4.** Quantitative analysis of supercoiled molecules. Density and Euler number were extracted from AFM images. A, B. Density and Euler number of highly –SC and +SC molecules on APTMS-treated surfaces (A) and in the presence of $Mg^{2+}$ ions (B). C, D. Density and Euler number values for all five plasmid populations obtained during RG reaction, on APTMS-treated surfaces (C) and in the presence of $Mg^{2+}$ ions (D). On APTMS, supercoiled molecules show higher densities, while more relaxed plasmids assume less dense shape. Similarly, the Euler number assumes the largest modulus for the most supercoiled forms, while the minimum modulus is obtained for relaxed plasmids. In contrast, deposition using $MgCl_2$ produces a monotonic trend in both density and Euler number: the –SC family (at 0 s) has a Euler number of -1 and low densities, while both Euler number modulus and density increase with increasing RG reaction time.

Taken together, these results suggest that, whereas the possibly 3D- shape of highly –SC plasmids is dependent on the deposition conditions, the structural configuration of +SC molecules is relatively insensitive to deposition conditions; in particular, highly +SC plasmids maintain their plectonemic arrangement under conditions that determine transition of –SC and intermediate topoisomers to looser geometry.

These results are the first experimental demonstration of the braiding effect induced by freely diffusing bivalent cations, and support a potential role of direct DNA–DNA interactions in tuning chromatin compaction [35, 6]. In fact, previous results obtained by molecular dynamics simulation demonstrated that the opposite chirality of superhelixes (positive for –SC and negative for +SC) produces different types of inter-segmental interactions along the plectonemes: +SC induces right-handed crossovers characterized by a strong groove-backbone interaction (i.e. the major grooves of one helix interlock with the minor groove of the other), while –SC supercoiling produces a simple juxtaposition of the grooves (i.e. major-major groove overlap) [37].

**2.2 AFM Imaging in liquid of –SC and +SC DNA molecules**

Our AFM images in air showed that only –SC molecules are heavily affected by the deposition procedure, while +SC shapes are definitely more stable. Since DNA may change its conformation during the steps of sample preparation and deposition, observation of molecules in solution may better reflect their conformation under physiological conditions and help overcome possible structure alterations arising from the rinsing and drying of the sample. Moreover, observation in liquid can be carried out under a number of controlled conditions. No data on +SC DNA in solution are available so far. We thus set up the experimental conditions for direct observation of –SC ($\sigma$=-0,04) and +SC ($\sigma \geq 0,04$) DNA molecules in aqueous solutions in the presence of mild ionic strength (100 mM NaCl). Under these conditions, the shapes of both –SC and +SC molecules looked very similar to those observed on APTMS in air (Fig. 5A and B).

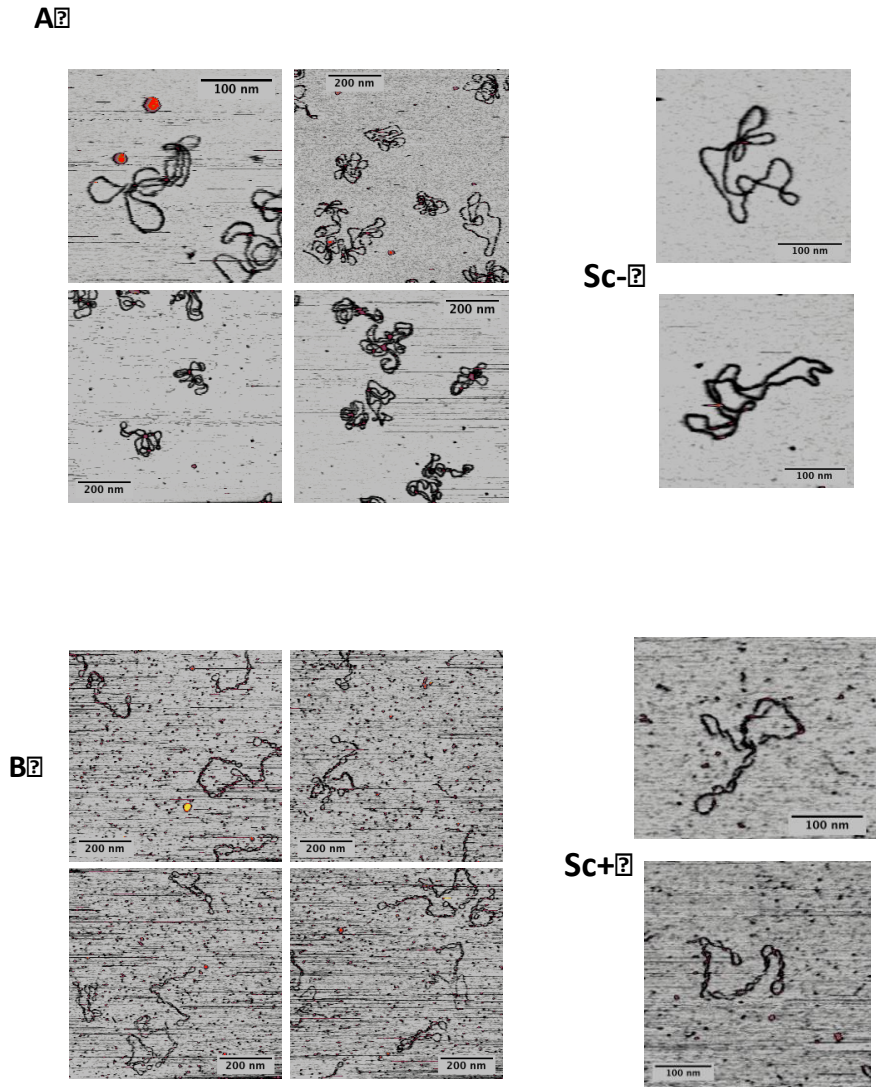

**Figure 5.** AFM analysis of supercoiled plasmids in solution. Representative images of –SC (A) and +SC (B) molecules at 35 °C are shown.

Assuming that both deposition methods permit molecular equilibration, the experimental data show that –SC , but not +SC plasmids were affected by the deposition conditions. The wounded (but not plectonemic) shapes assumed by –SC resemble the hyperplectonemes recently described [38] and can be the result of a negligible braiding effect: the distribution between Wr and Tw are mainly due to the mechanical stress induced by the supercoiling. On the other hand, +SC molecules exploit the synergistic effect of both steric and electrostatic interactions that favour braiding and development of plectonemes. Thus, despite the "mild" conditions used to deposit the plasmids, the incubation with $Mg^{2+}$ induces large modifications of the plasmids tertiary structure in –SC but not +SC molecules. This is a quite unexpected and novel result, since it demonstrate a clear asymmetric behaviour of plasmids with different SC sign.

## 2.3 Temperature effect on plasmids supercoiling

It is well known that temperature is an important parameter affecting DNA topology. However, the effect of temperature on supercoiling has been analysed by AFM only for –SC plasmids in air [39]; this study showed that increasing temperature from 25 to 50 °C induces a shift of the plasmid structure from supercoiled to more relaxed shapes. Moreover, observation at higher temperatures was accompanied by the appearance of globular structures, which became predominant at 80 °C [40] These structures were interpreted as a result of denaturation of localized regions, followed by collapse of the single strands. Finally, at 100 °C DNA was degraded into small fragments [40]. Since these experiments were performed in air, it is not known whether these structures might be due to deposition conditions. Moreover, no data on the effect of high temperature on +SC DNA are available.

Since positive supercoiling has been associated with life at high temperature, we sought to set up a protocol where plasmids were incubated at a given temperature and then directly observed in solution at the same temperature. As shown in Fig. 6A, after incubation for 10 min at 90 °C –SC molecules assumed globular and aggregated structures, confirming previous observations in dried samples [40]. In striking contrast, positive molecules conserved their plectonemic structure almost intact after the same treatment (Fig. 6B) (for comparison purposes AFM images of the same pool of plasmids acquired at 35 °C are reported in Fig. 1S).

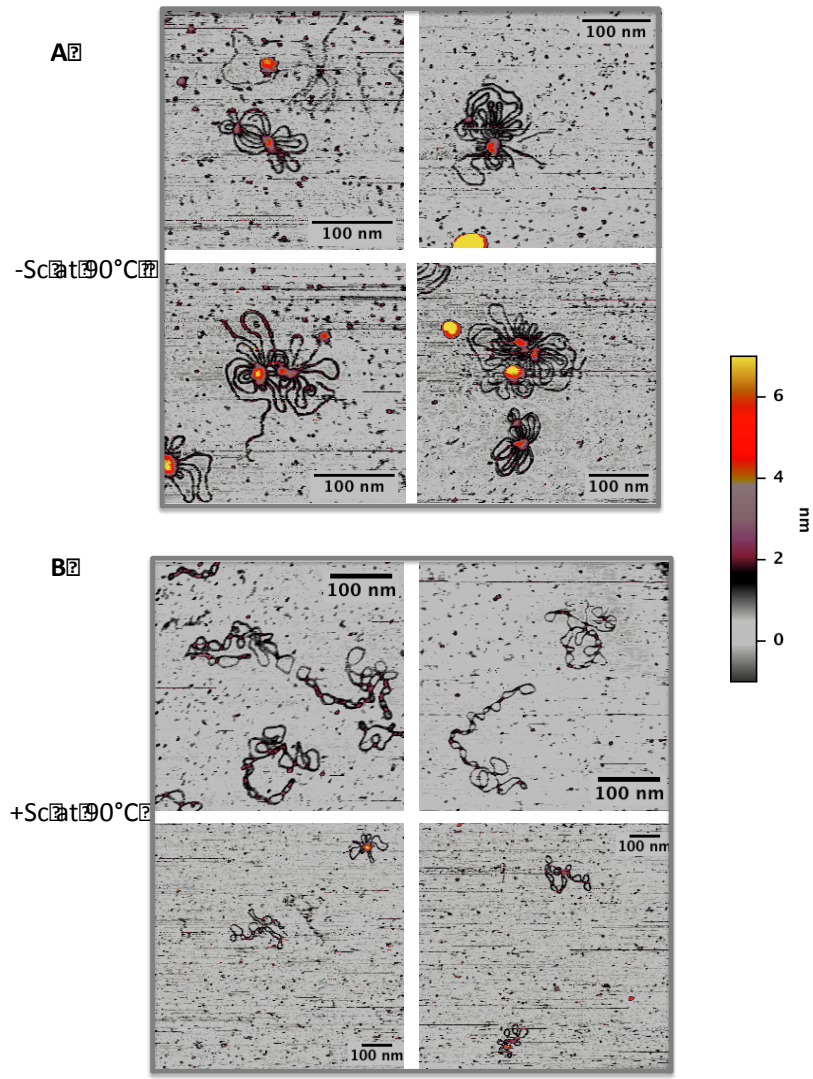

**Figure 6**. AFM analysis of supercoiled plasmids in solution at high temperature A) –SC + in liquid at 90 °C. Plasmids lost their plectoneme shape and assume a globular form compatible with denaturation of their structure. B) +SC plasmid in liquid at 90 °C. Nearly all plasmids maintain their plectonemic shape, thus confirming a more stable behavior of these molecules against thermal denaturation.

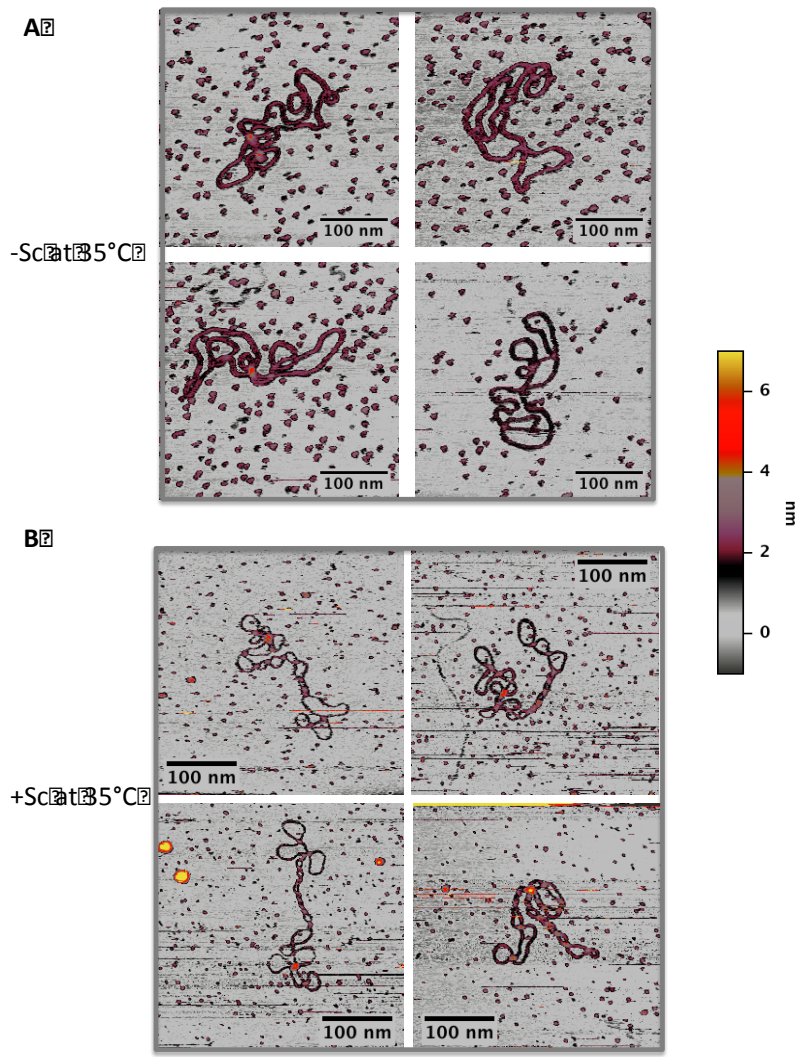

**Figure 1S.** AFM analysis in solution at 25°C. Representative conformations of –SC (A) and +SC (B) molecules.

Previous experiments showed that, whereas –SC molecules (minicircles as well as plasmids) expose single strand regions even at room temperature, +SC molecules do not [27,28]. We set up a similar experiment with our molecule populations: when incubated with the ssDNA endo-exonuclease Bal31 (which is able to cut single strand DNA in denatured bubbles), –SC DNA was degraded completely after 1 hour at 30°C (Fig. 7A, lane 1-4), whereas highly +SC plasmids (σ ≥ +0,04) were resistant to Bal31 digestion, confirming that they do not expose single strand regions at this temperature (Fig. 7A, lane 5-8).

Since our AFM analysis showed a striking structural stability of +SC molecules even at 90°C, we sought to find an alternative biochemical tool to probe the 2D structure of our molecules at high temperature. To this aim we used glyoxal, a small molecule that traps exposed bases and is resistant to high temperature. When incubated with 1 M glyoxal at room temperature, the –SC DNA band was shifted, indicating binding of the drug to single strand regions (Fig. 7B, lane 1-4); in contrast, the mobility of the +SC DNA was not affected when incubated with increasing concentrations of glyoxal (Fig. 7B, lane 5-8). In order to evaluate the effect of temperature, –SC and +SC DNA molecules were incubated for 5' minutes at increasing temperatures (from 80°C up to 90°C) in the presence of glyoxal. As expected, for the underwound DNA the glyoxal-induced shift increased with temperature (Fig. 7C, lane 1-3), because temperature increase further induces local denaturation. Interestingly, +SC plasmids did not bind glyoxal at any temperature up to 90°C (Fig. 7C, lane 4-6) indicating that they do not contain unpaired bases when incubated for 5 min up to 90 °C and are highly resistant to thermal denaturation.

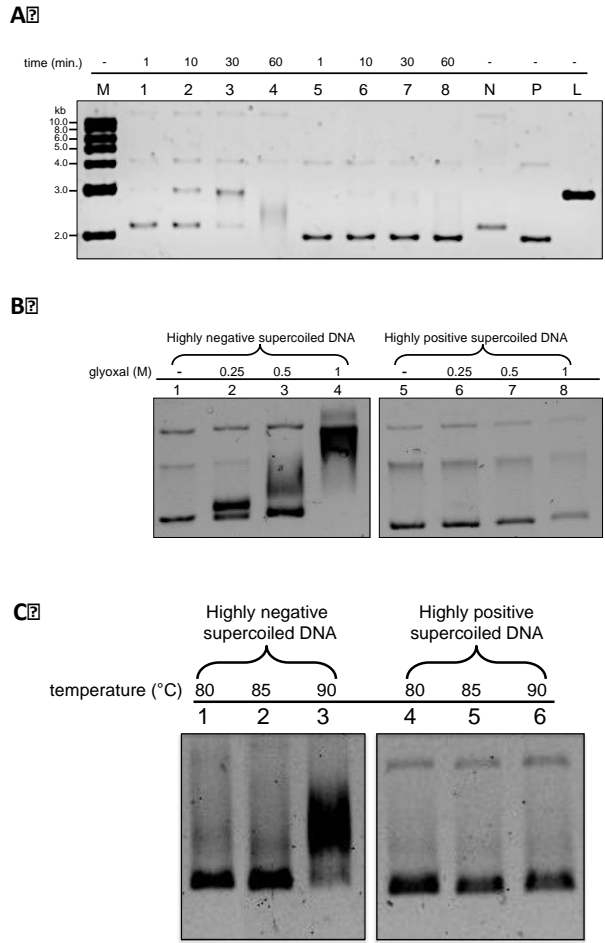

**Figure 7.** Biochemical analysis of supercoiled plasmids A. Plasmid DNA preparations were incubated with nuclease Bal-31 for increasing timespans, as indicated: lanes 1-4, –SC plasmid; lanes 5-8, +SC plasmid . N: mock-incubated –SC DNA; P: mock-incubated highly +SC DNA, L: linearized 3000 bp plasmid; M: 1 kb DNA ladder. Over time, samples were removed, quenched by the addition of stop buffer and the products analysed on 1% agarose gel. B. Incubation of plasmid DNA with glyoxal. Highly –SC and +Sc DNA preparations were incubated overnight in the absence or presence of increasing concentrations of glyoxal, as indicated, and analyzed on 0.8% agarose gel. Lanes 1-4, –SC DNA; lanes 5-8, +SC DNA. C. Effect of temperature on glyoxal binding. Highly –SC and +SC DNA were incubated for 5 min. at indicated temperatures in the presence of 1 M glyoxal and analyzed by 0.8% agarose gel electrophoresis.

In a similar analysis performed on DNA minicircles, Irobalieva and colleagues found that topoisomers with ΔLk ranging from -6 to -1 showed exposed bases, whereas those with ΔLk= 0, +1 and +2 did not, in line with our results. However, they also found that the +3 topoisomer was sensitive to Bal31 degradation, albeit at lesser extent as compared with –SC DNA molecules. This result was explained assuming that overwinding induces a sharp bending in this small topoisomer, resulting in local denaturation [27]. In contrast, we could not observe any evidence of denaturation in our +SC with very high ΔLk (≥+12), suggesting that these larger molecules can accommodate the Wr excess without the need of bending or denaturation to compensate for the torsional stress.

3. Conclusion

In this work we exploited AFM to perform for the first time a systematic analysis of plasmids having excess linking number between -12 to ≥+12 ($|\sigma|\approx$ 0.04), covering the full range of physiologically relevant topological states. By exploiting the high resolution imaging of DNA we studied how plasmid conformation is affected by immobilization, surface and temperature. We found a strong asymmetry in the behaviour of –SC and +SC molecules (summarized in Fig. 8 A). Both –SC and +SC molecules deposited and dried onto APTMS-treated surfaces, assumed plectonemic shapes. In contrast, the presence of cations ($MgCl_2$) during deposition influenced the shape of –SC molecules, which assumed an open conformation, but not of +SC molecules, suggesting that overwinding protects DNA from cations-dependent alterations. Moreover, direct observations in liquid showed a remarkable structural stability of +SC molecules at high temperature, whereas –SC molecules collapsed above 90 °C. These data, confirmed by biochemical analysis, are in line with the biological role of both negative supercoiling, which is thought to facilitate breathing of the double helix, and positive supercoiling, which is likely to contribute to DNA stability under extreme conditions.

Showing that the opposite signs of DNA affect the physical properties of the superhelices, our study provides new insights that may contribute to understand the role of specific DNA topology in the cellular environment as well as the physiological function of reverse gyrase and its hallmark reaction. Our results also show that AFM is a powerful tool to correlate different shapes of DNA molecules to different topologies, upon deposition in the presence of $MgCl_2$. For instance, by AFM it is possible to analyze, identify and quantify at single molecule level: i) DNA samples with mixed topologies; ii) different local topologies in DNA-protein complexes: iii) supercoiling-dependent binding of specific proteins (Fig. 8B). Our results might also be useful to explore DNA dynamics and its interactions in confined space [43] with other charged molecules, such as chemotherapeutic compounds, intercalating agents, and proteins involved in DNA metabolism.

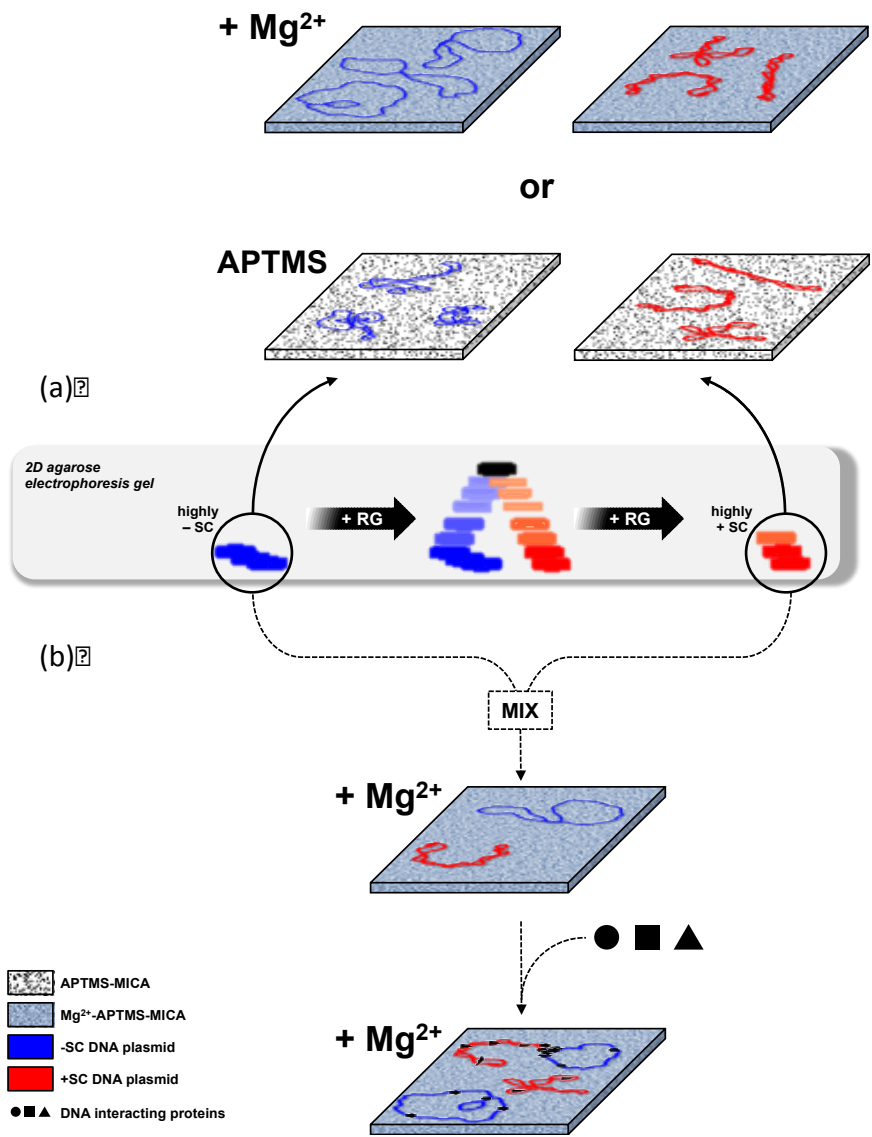

**Figure 8:** Schematic representation of –SC and +SC plasmids analysed by AFM: results obtained in this study (**A**) and their potential application to study DNA topology and its interactions with other molecules upon MgCl$_2$ deposition (**B**).

## 4. Experimental Section

### 4.1 RG time course reaction

For time course experiments, we incubated the RG from *S. solfataricus* [30]) (80 nM) with –SC pBs plasmid (10 nM) in a final volume of 150 µL of 1× RG buffer (35 mM Tris-HCl, pH 7.0, 0.1 mM Na$_2$EDTA, 30 mM MgCl$_2$, 2.0 mM DTT, 1 mM ATP). Reactions were incubated at 90 °C and withdrawals of 30 µL were done at different times as indicated. After incubation, DNA samples were purified by phenol extraction and ethanol precipitation; an aliquot (300 ng) of each sample was analysed by two-dimensional gel electrophoresis in 1.2% agarose and 1× Tris borate-EDTA buffer [41]. Gels were stained with ethidium bromide (1 µg ml$^{-1}$) and destained in deionized water for the analyses with UV light by a VersaDoc 4000™ apparatus and the

QuantityOne software (Bio-Rad). Another aliquot of the same samples was used for AFM analyses.

## 4.2 Determining supercoiling density of plasmid DNA molecules

Topological properties of DNA are defined by few parameters: twist ($T_w$, the number of times each helix twists around the other) and writhe ($W_r$, the number of crossings the double helix makes around itself); in a covalently closed DNA molecule, the sum of these two parameters is a topological invariant, called linking number ($L_k=T_w+W_r$). Because the energy required to supercoil a polymer depends on its length, a useful parameter to compare different topological isomers (topoisomers) is the supecoiling density $\sigma=(L_k-L_{k0})/L_{k0}$, where $Lk^0$ and $Lk$ represent the DNA linking number for the relaxed and the supercoiled DNA, respectively. The superhelical density of topoisomers with different supercoiling degree was estimated by 2D gel electrophoresis. Briefly, after migration the gel was stained with ethidium bromide (1 µg ml$^{-1}$), destained, and photographed under UV light. The relative intensity of each band was measured and the DNA linking number change (ΔLk) of corresponding topoisomers was determined and used to obtain σ values applying the above equation. The σ values of all topoisomers whose intensity was >30% of the intensity of the most abundant topoisomer were used for the calculation of the mean σ value.

To increase the resolution of highly negatively supercoiled molecules, which run as single bands in 2D gels, samples was subjected to monodimensional electrophoresis in the presence of a small concentration of the intercalating agent chloroquine, which causes a decrease in twist (Tw) and consequently an increase of writhe (Wr). Under these conditions, the band of the negative plasmid was split into a few bands, with ΔLk ranging from -7 to -12. The σ of the most –SC topoisomer (ΔLk=-12) is -0.04, in line with the supercoiling density of plasmids extracted from *E. coli* cells cultures [42] (Fig. 2S).

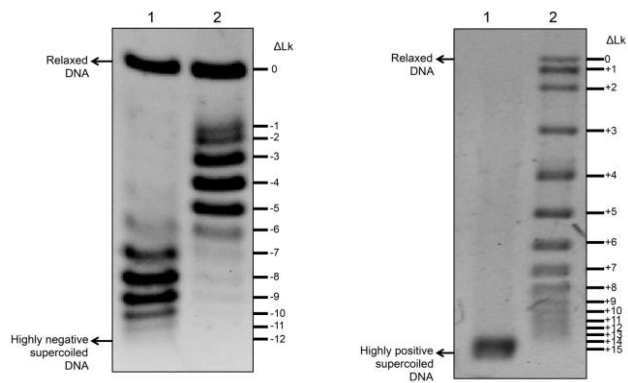

**Figure 2S. Determining supercoiling density of –SC and +SC plasmid. A.** Gel electrophoresis run in the presence of 1 µg mL$^{-1}$ chloroquine to determine the superhelical density of negatively supercoiled plasmid. Lane *1,* plasmid extracted from *E. coli* cells grown at 37 °C; lane *2,* ladder of negative topoisomers with the indicated linking number. **B**. Gel electrophoresis without intercalating agents to determine the superhelical density of the end point product of the reverse gyrase; reaction was at 90 °C for 10 min in the presence of 1 mM ATP, with the plasmid substrate shown in lane 1 of fig 1A. Lane 1*,* positively supercoiled plasmid; lane 2*,* ladder of positive topoisomers of the indicated ΔLk. Arrows indicate the relaxed and supercoiled forms of the plasmid, respectively. The calculated sigma (σ) values are reported.

For the +SC topoisomer population, monodimensional electrophoresis without intercalants was used, showing that they still migrated as a single band, whose σ was ≥ 0.04 and it was not possible to determine whether this population is comprised of a single or a few compact topoisomers.

### 4.3 DNA deposition protocol

For all samples, DNA stock solution was mixed with ultrapure water, Tris buffer at pH 7.5 (final concentration 7 mM) and salts (see below), achieving a final DNA concentration of 2 ng µL-1. To induce a reproducible and attractive interaction between the plasmids and the surface of the mica we either added $MgCl_2$ (3 mM final concentration) to the solution or we employed APTMS-treated-mica. In this latter case NaCl at a final concentration of 100 mM was added to the mixture. For measurements in air, plasmids were incubated for 3 min adding 20 µL of the final solution to the mica, afterward the sample was gently rinsed with 0.6 ml of ultrapure water and dried under a mild air flow.

For measurements in liquid, two different procedures were used, both exploiting AP-mica as substrate. For measurements at low temperature (35°C), NaCl was added to the mixture at a final concentration of 100 mM, incubating 20 µL of the final solution on mica for 3 minutes. Then, samples were gently rinsed adding and removing 80 µL of the buffer solution used to dissolve the plasmids. For plasmid DNA high temperature measurements, the following procedure was used: first the plasmid solution was heated at 90°C for 10 minutes, then 25 µL of this solution were incubated on the APMTS-treated mica substrate at 90°C for 3 minutes, supplying buffer to compensate for the evaporation and to avoid concentration effects. Finally, the incubated mica was gently rinsed (at room temperature) adding and removing 80 µL of the same buffer solution used to dissolve the plasmids. No salt was added in this case, to increase plasmid affinity toward mica at this high incubation temperature. In both cases, 100 µL of the corresponding buffer were added to the sample before insertion in the AFM cell.

APTMS treated mica was obtained exposing freshly cleaved mica substrates to APTMS vapors in vacuum for 3 minutes and used immediately afterward.

### 4.4 AFM Imaging

AFM images were collected using an Asylum Research Cypher equipped with an environmental scanner. Olympus AC200TS and BL-AC40 cantilevers were used in tapping mode in air and liquid, respectively. Samples were scanned at a rate of about 1-2 Hz and images were obtained at several separate mica locations. All the AFM acquisitions in air were performed at room temperature, while acquisitions in liquid were performed at 35°C. Asylum provided software was use d for data acquisition. AFM visualization was performed using ImageJ [26] and dedicated plug-ins that we developed.

4.5 **Nuclease Bal-31 assay**

–SC and +SC plasmid DNA (400 ng in 40 µL final volume) was incubated with 0.2 units of nuclease Bal-31 at 30 °C in 20mM Tris-HCl, pH 8.0, 600 mM NaCl, 12 mM $MgCl_2$, 12mM $CaCl_2$ and 1 mM disodium EDTA. At 1, 10, 30 and 60-minute intervals, 10 µL (100 ng) samples were removed, stopped by addition of an equal volume of stop buffer (50 mM Tris-HCl, pH 8.0, 100 mM disodium EDTA, 10% glycerol, 200 µg ml$^{-1}$ proteinase K), followed by incubation at 45 °C for 30 min to degrade Bal-31. Products were analyzed by electrophoresis through 1.2% agarose in 1× Tris-Borate EDTA. Gels were stained with ethidium bromide (1 µg ml$^{-1}$) and destained in deionized water for the analyses with UV light by a Bio-Rad VersaDoc apparatus.

**4.6 Glyoxal assay**

Glyoxal was first deionized with AG-501-X8 mixed bed ion-exchange resin (Bio-Rad, Hercules, CA). –SC and +SC plasmid DNA (100 ng) was incubated with increasing concentrations of glyoxal (from 0.25 to 1 M) in 10 mM sodium phosphate, pH 7.0, for 16 h at room temperature. Control reactions were incubated in 10 mM sodium phosphate only. Samples were analysed by electrophoresis through 0.8 % agarose gel in 1× Tris-Borate EDTA. Gel was analysed as already reported.

In order to determine the effect of temperature on glyoxal binding, we performed the reaction by incubating 100 ng of negatively or positively supercoiled DNA in presence of 1 M glyoxal for 5 min. at indicated temperatures. After incubation, samples were analysed as described above.